\documentclass[11pt, A4paper]{article}

\usepackage{amsmath,amsthm,amssymb,amsfonts}
\usepackage{graphics,graphicx}
\usepackage{epsfig}
\makeatletter
\@addtoreset{equation}{section}
\makeatother

\begin{document}

\begin{center}
\Large\textbf{Non-relativistic Matrix Inflation}
\end{center}
\vspace{0.3cm}
\begin{center}
\large{\bf Aaron Berndsen} ${}^{a,}$\footnote{aberndsen@sfu.ca} \large{\bf James E. Lidsey} ${}^{b,}$\footnote{j.e.lidsey@qmul.ac.uk} \large{ \bf and John Ward} ${}^{c,}$\footnote{jwa@uvic.ca} \\
\vspace{0.7cm}
\emph{${}^a$ Physics Department, Simon Fraser University \\ Vancouver, BC \\ V5A 1S6, Canada \\}
\vspace{0.4cm}
\emph{${}^b$ Astronomy Unit, School of Mathematical Sciences \\ Queen Mary University of London \\ London, E1 4NS, UK \\}
\vspace{0.4cm}
\emph{ ${}^c$ Department of Physics and Astronomy, University of Victoria \\ Victoria, BC \\ V8P 1A1, Canada}
\end{center}
\vspace{1cm}
\begin{abstract}
We reconsider a string theoretic inflationary model, 
where inflation is driven by $n$ multiple coincident 
$D3$-branes in the finite $n$ limit. We 
show that the finite $n$ action can be continued to the limit of 
large $n$, where it converges to the action for a wrapped D5-brane
with $n$ units of $U(1)$ flux. 
This provides an important consistency check of the scenario 
and allows for more control over certain back-reaction effects. 
We determine the most general form of the action for a 
specific sub-class of models and examine the non-relativistic 
limits of the theory where the branes move at speeds much less than the
speed of light. The non-Abelian nature of the 
world-volume theory implies that the inflaton field is matrix valued
and this results in modifications to the 
slow-roll parameters and Hubble-flow equations. 
A specific small field model of inflation is investigated
where the branes move out of an AdS throat, and observational
constraints are employed to place bounds on the background  
fluxes.  
\end{abstract}
\newpage
\section{Introduction}

Modulo several signatures from the strong-coupling regime of 
QCD~\cite{Policastro:2001yc,Brower:2000rp,Luscher:2002qv}, as
well as indirect evidence possibly emerging from terrestrial
accelerators~\cite{Hewett:2005iw}, string theory remains best-tested in
the realm of early universe cosmology. Observations of the cosmic
microwave background (CMB) provide strong evidence for the fiduciary
$\Lambda$CDM model of cosmology, together with a period of primordial
inflation which set the initial conditions for the density fluctuations. 
Ideally primordial inflation occurred in an energy
regime where degrees of freedom unique to string theory are excited; or,
at the very least, string theory should provide the mechanism for an epoch of
inflation that is consistent with current observations. An
understanding of early-universe dynamics therefore provides a direct means of
exploring aspects of string theory. 

Given this motivation, inflationary model building within string theory
has become an established research direction, yielding many
different models. Despite the large number of proposals, this pursuit
serves---at the very least---to constrain  the parameter space and overall
viability of string theory~\cite{McAllister:2007bg}. A simple dichotomy
of these scenarios emerges from the origin of the inflaton, i.e., 
whether this field originates from the open or
closed sector of the theory. In both cases, numerous novel
phenomena can arise and, in this sense, string theory provides a natural
explanation for possible effects which can not be understood within the 
context of vanilla, slow-roll inflation. Since such results 
reside mostly within the framework of effective field theory,  
one cannot be certain if string
theory is the unique UV completion. Nonetheless,
it is remarkable that the classes of
models selected by string theory have such a wealth of 
unique and potentially detectable phenomena. 

In this paper we focus on a popular form of brane inflation embedded in the
open-string sector of string theory. This model of inflation
associates the inflaton with a modulus parametrizing the separation of a
probe brane from a stack of branes residing at the bottom of a warped
throat. There has been immense excitement that such a model could lead to 
distinct observational signatures, particularly regarding a non-vanishing 
bi-spectrum in the primordial curvature perturbation (see
\cite{RenauxPetel:2009sj, Mizuno:2009mv, Panotopoulos:2007ky,Smidt:2009ir} 
for examples). 
However, there remain many open questions regarding the regime of 
validity of this scenario, at both  the theoretical and observational
levels \cite{Chen:2008hz, Kobayashi:2007hm,Becker:2007ui}. 

In view of this, we go beyond the simplest approximation 
and consider a more complicated model comprised of $n$ multiple
 coincident branes that follow a single trajectory. 
Due to the nature of the world-volume theory living on these branes, the
resulting field-theoretic analysis is automatically a non-Abelian 
theory, where the inflaton is now 
a scalar field transforming in an adjoint representation. One strong 
objection to using a theory of this form is that
we wish the standard model to exist in another (warped) throat, and 
therefore a world-volume description of inflation
is not desirable. However, we will argue that such a theory has a dual 
description in terms of a wrapped higher-dimensional
brane, and therefore represents an important theory within 
the class of DBI-inflation 
models. Importantly, we will argue that 
the wrapped brane description is valid for large values of $U(1)$ flux 
by showing how the multi-brane action converges
as we take the large $n$ limit of the finite $n$ theory.

Since the fully relativistic theory has been (relatively) well explored
\cite{Huston:2008ku, Ward:2007gs,Thomas:2007sj,Alabidi:2008ej}, 
we focus in the present work on the non-relativistic limit,
where the velocity of the inflaton field is significantly 
smaller than the speed of light. It is in this sense that we refer to such a  
regime as `non-relativistic' and this is distinct from the limit 
where the effective sound speed of inflaton field fluctuations 
is much less than that of light. 
In the standard DBI-inflation model, the non-relativistic
limit of the theory is simply a canonical scalar field in the 
appropriate FRW-background. However, due to the non-commutative structure
inherited from the non-Abelian field theory, the non-relativistic limit 
in our scenario is a non-canonical scalar field theory. 
Consequently, one should anticipate different physical effects 
when compared to the standard `slow roll' inflationary scenario.

The paper is structured as follows. 
In section~\ref{sec:intro}, we remind the reader of the string theoretic
construction for this class of models. Firstly  
we argue that the large $n$ limit of the 
multi-brane theory is precisely dual to a single wrapped $D5$-brane with 
$n$-units of $U(1)$ flux.
We then consider the finite $n$ limit of the multi-brane theory, 
since this is analytically tractable, and proceed to 
argue that the finite $n$ limit does indeed analytically converge 
with the large $n$ theory in the correct scaling limit. This is the first time 
that such a result has been established in the literature.
In section \ref{sec:finiten}, we develop the framework for analyzing 
inflationary dynamics in the finite $n$ theory, and construct
a simple example of small-field inflation in an AdS background  
driven by an inverted harmonic oscillator potential. 
In section \ref{sec:twodim}, we specialize to
the $n=2$ theory and consider in some detail how the inflationary observables 
differ from those of the single brane model. We then 
discuss how the flow equations for such a model can be constructed. 
Finally, we conclude in section \ref{sec:discuss} 
with some general remarks and a discussion of future directions.


\section{The Action for matrix cosmology}\label{sec:intro}

\subsection{Matrix cosmology at large $n$}

Numerous extensions of the vanilla DBI-inflation model have been proposed, 
yielding novel results and phenomena. These include, but are not limited to,
multiple-field scenarios 
\cite{Langlois:2008wt,  Langlois:2008qf, Huang:2007hh, Easson:2007dh}, 
multiple branes \cite{Cai:2008if, Cai:2009hw, Cline:2005ty}, monodromies
\cite{Silverstein:2008sg}, Wilson lines 
\cite{Avgoustidis:2006zp, Avgoustidis:2008zu} and the inclusion of 
bulk form fields \cite{Langlois:2009ej}.
In this section, we are interested in the more conservative 
generalization of the model, where the 
solitary $D3$-brane is replaced by a solitary $D5$-brane 
wrapping a non-trivial two-cycle of the
internal geometry  (usually taken to be an $S^2$---note that this is a simply-connected space and therefore we must turn on
some world-volume flux to stabilize the brane)  
\cite{Kobayashi:2007hm, Becker:2007ui}. 
Whilst there has been significant progress in the field of flux 
compactification in the type II
theory onto (conformal) Calabi-Yau manifolds, we will consider 
a simpler class of warped background geometries 
based upon the well understood conifold geometry
\begin{equation}
ds^2 = h^{2}(\rho) ds_4^2 + h^{-2}(\rho)(d\rho^2 + \rho^2 d\Omega_{X_5}^2).
\end{equation}
The basic idea is that fluxes in the compactification back-react 
to form a throat (parametrized by a radial coordinate $\rho$)
over the base manifold, $X_5$.
We allow the brane to be localized (and flat) in the large dimensions, 
wrapping a two-cycle within the internal manifold. We further assume that 
the brane is dynamical along the throat direction and freeze out any angular 
degrees of freedom. 

Without specializing to a particular supergravity background solution for 
the warp factor, $h(\rho)$, we know that generically one must turn on 
magnetic flux along the two-cycle 
directions in order to stabilize the configuration. The 
$U(1)$ field strength must be proportional to the volume of the wrapped cycle,
where the constant of proportionality is determined by the $n$ units of flux 
threaded through the $S^2$: 
\begin{equation}
F^{(2)} = \frac{n}{2} \omega_2 , 
\end{equation}
where $\omega_2$ is the associated two-form of the sphere.

The corresponding action for a $D5$-brane is then given by the usual 
DBI expression (see \cite{Myers:1999ps} for further discussion): 
\begin{equation}
\label{ac}
S = -T_5 \int d^6 x e^{-\Phi} \sqrt{-\rm{det}(\hat{G}_{ab}+\hat{B}_{ab}+\lambda
F_{ab})} - \mu_5 \int \left(\sum_n \hat{C}^{(n)}e^{\hat{B}} \right) e^{\lambda F} ,
\end{equation} 
where a $\hat{} $ denotes the pullback of that particular 
tensor to the world-volume, and $\lambda$ is the inverse of the fundamental 
string tension which couples to the $U(1)$ field strength. 
The field $\Phi$ is the dilaton defining the string coupling constant in the
low energy theory. However, since we are assuming that the throat is sourced by 
$D3$-brane charge, we can set this term to unity without loss of 
generality. For more involved backgrounds, 
the dilaton will indeed be non-trivial and can lead to more 
complicated dynamics.

The second term in Eq. (\ref{ac}) corresponds to the coupling of the
brane to the bulk form fields in the $RR$-sector once they are 
pulled-back to the world-volume. The presence of the summation 
indicates that there can be
coupling to form fields of lower degree, provided that there 
is a non-zero $B$ (or $F$) term. Note that $T_5$ and $\mu_5$ denote,
respectively, the tension and charge carried by the brane, and that
these are related via supersymmetry. By utilizing the above metric and
gauge field ansatz we can compute the action for the wrapped
$D5$-brane. It is given by    
\begin{equation}\label{eq:D5action}
S = -4\pi T_5 \int d^4 \xi \left(h^2 \sqrt{1-h^{-4}\dot{\rho}^2}
\sqrt{\frac{1}{4} h^{4} \lambda^2 n^2 + \rho^4} -h^4 \frac{n}{2}\lambda \right) ,
\end{equation}
where we have worked in physical coordinates 
and integrated out the directions along the $S^2$. 
For cosmologically relevant solutions one must  
minimally couple this term to the usual Einstein-Hilbert action. 
For a concrete embedding of this action into a particular string theory
background we refer the reader to \cite{Becker:2007ui}.

One of the many interesting features of string theory is 
the existence of duality symmetries, 
whereby a brane configuration can be related to another (equivalent) one. In
this instance, we can employ our knowledge of the Myers dielectric effect 
\cite{Myers:1999ps} to understand how the above action is captured in 
the dual picture by $n$ coincident $D3$-branes 
expanding along a fuzzy two-sphere\footnote{See the alternate 
proposal by \cite{Tseytlin:1997csa}.}.
The full discussion is intricate and chronicled at length elsewhere
(see \cite{Thomas:2006ac, Iengo:2007cp} for examples in non-trivial
backgrounds). We therefore quote the results and refer the reader to
\cite{Ward:2007gs, Thomas:2007sj} for more details on the relation to
cosmological model building. The validity of the Myers action using
boundary fermions has been considered at length in \cite{Howe:2006rv,
  Howe:2007eb}, and can certainly be trusted at leading order. An
important related usage of the action is discussed in
\cite{Brown:2009yb}. 

The action for $n$ coincident branes in the 
limit where $n \gg 1$ can be written as
\begin{equation}\label{eq:nD3action}
S = - n T_3 \int d^4 \xi h^4 \left(\sqrt{1-h^{-4} \dot{\rho}^2}\sqrt{1+\frac{4 \rho^4h^{-4}}{\lambda^2 n^2}}-1 \right)\,,
\end{equation}
and it can be shown that this is exactly the same 
as the action (\ref{eq:D5action}) 
above\footnote{At least to leading order. Examining the duality to higher 
orders is certainly interesting, but highly non-trivial.} 
by employing the known relationship
between the brane tensions, $T_3 = 4 \pi^2 \alpha' T_5$, and  
by identifying the large $n$ limit in both cases.  
This implies that in the (macroscopic) $D5$-brane theory, we must take 
the flux to be large. Since our background is relatively simple, one 
could also consider the S-dual theory, where the $D3$-branes expand into a 
wrapped NS$5$-brane. In this case, the question of 
where the standard model degrees of freedom reside becomes crucial
since fundamental strings cannot end on the 
world-volume of the fivebrane. Within this context, a fivebrane model was 
recently proposed as
a concrete realization of axionic monodromy inflation \cite{Flauger:2009ab}.

An important result 
is that the back-reaction of higher-dimensional branes becomes 
important in the relativistic regime \cite{Kobayashi:2007hm}. 
Given our dual interpretation of the $D5$-brane action, this is 
intuitively obvious, since we are essentially taking $n\gg1$, and this 
will typically back-react on the geometry and invalidate the
probe-brane approximation. Thus, it is clearly desirable to 
construct a theory where $n$ is not taken to be extremely large. 
This should reduce the
back-reaction and allow for more control over the theory. 
However, constructing the action for multiple $D3$-branes in the finite $n$ 
limit is a highly technical issue and has only been 
resolved in a few simple cases. Nonetheless, 
progress has been made and in the following subsection, the
key results that are relevant to the present discussion 
are summarized. Full details of the construction can be found  
in \cite{Huston:2008ku}. 


\subsection{Matrix cosmology at finite $n$}

The tractability of the problem for finite $n$ is intimately tied to the 
symmetrized trace prescription (${\rm STr}$) associated with open string scattering 
amplitudes. This suggests that one must first average over all symmetric 
permutations before taking the gauge trace.
Fortunately, one can make headway by identifying the scalar fields with a 
finite dimensional irrep of $SO(3) \sim SU(2)$,  
which leads to the following result \cite{Ramgoolam:2004gw, McNamara:2005ry}:
\begin{eqnarray}
{\rm STr}(\alpha^i \alpha^i)^q &=& 2(2q+1)\sum_{i=1}^{n/2}(2i-1)^{2q} \hspace{0.7cm}  n \hspace{0.2cm}\rm{even} \nonumber \\
&=& 2(2q+1)\sum_{i=1}^{(n-1)/2}(2i)^{2q} \hspace{0.8cm} n \hspace{0.2cm}\rm{odd}\,.
\end{eqnarray}
As an aside we remark that for the flat $D3$-branes, 
the scalar fields parametrize the direction of a transverse
$S^2$---assuming that the full, non-compact, background geometry admits a
sub-manifold with an $SO(3)$ isometry. The corresponding action for
$n$ multiple $D3$-branes is then given by the usual expansion of the
Myers DBI action:  
\begin{equation}
P_n = -T_3 {\rm STr} \left(h^4(\rho) \sum_{k,p=0}^{\infty}(-Z\dot{R}^2)^k Y^p (\alpha^i \alpha^i)^{k+p} {1/2 \choose k}{1/2 \choose p}+V(\rho){\bf 1}_n - h^4(\rho){\bf 1}_n \right)
\end{equation}
where we have defined the following terms
\begin{equation}
\label{defZ}
Z = \lambda^2 h^{-4}(\rho) , 
\hspace{0.5cm} Y = 4 \lambda^2 R^4 h^{-4}(\rho) , 
\hspace{0.5cm} {1/2 \choose q} = \frac{\Gamma(3/2)}{\Gamma(3/2-q)\Gamma(1+q)}
\end{equation}
and the fuzzy sphere radius $R$ is related to the 
physical radius $\rho$ and the number of branes via
\begin{equation}
\rho^2 = \lambda^2 R^2 (n-1)^2.
\end{equation}
Those familiar with such a construction should 
note that one usually relates the physical
coordinate to the quadratic Casimir $(n^2-1)$. 
However, for finite $n$ there is a more precise
definition of the physical radius in terms of ratios of 
operators \cite{McNamara:2005ry}. This is a necessary
requirement for the convergence of the theory, and agrees with 
the definition in the limit of large $n$.
It should be noted that the radial coordinate $\rho$, 
the potential $V(\rho)$, and the warp factor $h(\rho)$ are all singlets 
under the symmetrized trace prescription. Moreover, the potential as
written here is dimensionless, since we have absorbed a factor of the brane 
tension into its definition. 
It is also important to note that the physical radius is a function 
of $n$ in this model and 
this will play a role when the recursion relations 
are employed. The key point is that
the scalar field, which we wish to identify with the inflaton, 
is now matrix valued\footnote{For recent additional work using the 
same idea we refer the 
interested reader to \cite{Ashoorioon:2009wa}.} and is in a finite 
dimensional representation of the corresponding gauge group.

The most important representation of $SU(2)$ is the fundamental 
(or two dimensional) one, since the representation is simply that 
of the Pauli spin matrices. It also turns out that the form of the symmetrized 
trace ansatz leads to a recursive structure for the theory, 
where we can construct
the action for any $n$ once the action for $n=2$ has been computed. 
Thus, the correct way to write the action is in terms of spin-$\frac{1}{2}$ variables as
first elucidated in \cite{Huston:2008ku}. It is found that the action 
takes the form  
\begin{equation}
\label{eq:recursive_action}
\fbox{$\displaystyle P_n(Z,Y)=\sum_{k=1}^{(n-1+\delta_n)/2}P_2((2k-\delta_n)^2Z,(2k-\delta_n)^2Y)-nT_3(V(\rho)-h^4(\rho))$}
\end{equation}
where $\delta_n = 1$ when $n$ is even and $\delta_n=0$ when $n$ is odd. 
A similar recursive structure exists for the energy density, 
$E_n (Z,Y)$ (more specifically the time-time
component of the kinetic part of the energy-momentum tensor).

For completeness, we write down the recursion functions $P_2$ and $E_2$: 
\begin{eqnarray}
P_2(Z,Y) &=& - \frac{2 T_3 h^4}{\sqrt{1+Y}}\left( \frac{1+2Y-(2+3Y)Z\dot{R}^2}{\sqrt{1-Z\dot{R}^2}}\right) \nonumber \\
E_2(Z,Y) &=& \frac{2 T_3 h^4}{\sqrt{1+Y}} \left(\frac{1+2Y-YZ\dot{R}^2}{(1-Z \dot{R}^2)^{3/2}} \right)
\end{eqnarray}
where $Z, Y$ are the functions defined in Eq. (\ref{defZ}). 
Note that the algebraic forms of these functions are significantly different 
from those of the $n=1$ action and, consequently, one should anticipate 
different physics to emerge. 


\subsection{Convergence with the large $n$ limit}

A crucial question that now arises is whether 
the finite $n$ action defined in Eq. (\ref{eq:recursive_action})
really does reproduce the action (\ref{eq:nD3action}) in the large $n$ limit. 
We present here, for the first time, an argument which suggests that this is 
indeed the case. Our approach is to expand both actions as Taylor series and
compare coefficients term by term. 
We focus on the $NS$-$NS$ sector, since
the $RR$-sector converges in a trivial manner. 
To proceed, we rewrite action (\ref{eq:nD3action}) 
at large $n$ in the more compact form
\begin{equation}
\label{rewriteaction}
P = -nT_3 h^4 \left(1+\frac{Y_0}{\hat{C}} \right)^{1/2} \sqrt{1-(n-1)^2 \dot{\psi^2}}
\end{equation}
where we have introduced the following variables:  
\begin{equation}
\label{defY0}
Y_0 = \frac{m_s^4}{\pi^2 T_3^2} \left(\frac{\phi}{h} \right)^4, \hspace{0.5cm} \dot{\psi}^2 = \frac{1}{(n-1)^2}\frac{\dot{\phi}^2}{T_3 h^4}, \hspace{0.5cm} \hat{C}=n^2-1,
\hspace{0.5cm} Y = \frac{Y_0}{(n-1)^4}
\end{equation}
The canonical inflaton field is defined in terms of the throat 
geometry via the standard relation $\phi = \sqrt{T_3} \rho$.

Expanding the DBI action requires the introduction 
of the binomial coefficient, which we write as
\begin{equation}
(1+x)^{\alpha} = \sum_{j=0}^{\infty} {\alpha \choose j}x^j,  \hspace{1cm} {\alpha \choose j} = \frac{\alpha(\alpha-1)\ldots(\alpha-j+1)}{j!}
\end{equation}
For the remainder of this subsection we explicitly drop the summation sign, 
although the summation over Latin indices is always implied.
The key point is to observe that the ratio $Y_0/\hat{C}$ 
will generally be small in the large $n$ limit. This implies that  
we can expand the action (\ref{rewriteaction}) up to the 
leading order terms in $(\dot{\psi}^2)^j Y^i$. It follows, upon expansion, that 
\begin{equation}\label{eq:action_at_large_n}
\fbox{ $\displaystyle P \simeq -\frac{n(n-1)^{2j}T_3 h^4}{4 (n^2-1)^i i! j!}\left\lbrack \left(-\frac{1}{2}\right) \ldots \left(\frac{3-2i}{2}\right)Y_0^i\right\rbrack
\left\lbrack \left(-\frac{1}{2} \right)\ldots \left(\frac{3-2j}{2} \right)(-\dot{\psi}^2)^j\right\rbrack.$}
\end{equation}
The action (\ref{eq:action_at_large_n}) is what we aim to reconstruct 
using the finite $n$ formalism. We demonstrate this explicitly for odd $n$ 
(the case of even $n$ is analogous). The relevant 
$n=2$ term, written in terms of the new variables (\ref{defY0}), is given by 
\begin{equation}
P_2 = -\frac{2 T_3 h^4}{\sqrt{1+Y}}\left(1+2Y-(2+3Y)\dot{\psi}^2 \right)\left(1-\dot{\psi}^2 \right)^{-1/2}.
\end{equation}
We now proceed to expand the velocity factor 
$(1-\dot{\psi}^2)^{-1/2}$
up to terms of order $(\dot{\psi}^2)^j$. After some algebra, we find that 
the relevant term in the expansion is given by 
\begin{eqnarray}
P_2 &\simeq & -\frac{2 T_3 h^4}{\sqrt{1+Y}}\left(1+2Y-(2+3Y)\dot{\psi}^2 \right)\left\lbrack \frac{-(\frac{1}{2}) \ldots (\frac{1-2j}{2})(-\dot{\psi}^2)^j}{j!} + \frac{-(\frac{1}{2})\ldots (\frac{3-2j}{2})(-\dot{\psi}^2)^{j-1}}{(j-1)!} \right\rbrack \nonumber \\
&\simeq& -\frac{2 T_3 h^4}{\sqrt{1+Y}}\left\lbrack \frac{-(\frac{1}{2}) \ldots (\frac{3-2j}{2})(-\dot{\psi}^2)^j}{(j-1)!}\right\rbrack \left\lbrack \frac{(1+2j)}{2j}+\frac{(2+2j)Y}{2j}\right\rbrack + \ldots.
\end{eqnarray}
Performing the analogous expansion on the $(1+Y)^{-1/2}$ factor
up to terms of order $Y^i$ then yields the result: 
\begin{eqnarray}
P_2 &\simeq& - \frac{2T_3 h^4(1+2j+2i)}{4i! j!}\left\lbrack \left(-\frac{1}{2}\right) \ldots \left(\frac{3-2i}{2}\right)Y^i\right\rbrack
\left\lbrack \left(-\frac{1}{2} \right)\ldots \left(\frac{3-2j}{2} \right)(-\dot{\psi}^2)^j\right\rbrack \nonumber \\
&\simeq& -2 T_3 h^4 \left(\frac{1+2j+2i}{4} \right)Q_Y Q_{\psi} \ldots \, .
\end{eqnarray}
However, we should recall that this is simply the expansion of the 
$n=2$ action and we must employ the recursion relations 
(\ref{eq:recursive_action}) in order to generate the full
structure for any value of $n$. For odd $n$, this is achieved 
through the rescaling 
\begin{equation}
\dot{\psi}^2 \to (2k)^2 \dot{\psi}^2, \hspace{0.5cm} Y \to (2k)^2 Y
\end{equation}
Furthermore, we must also introduce a new summation over the $k$ 
variable to obtain the complete expression for the 
leading order terms in $(\dot{\psi}^2)^j Y^i$ in the action $P_n$.  
Taking these considerations into account, we arrive at the expression
\begin{equation}
\label{actionsofar}
P_n \simeq -2 T_3 h^4 Q_Y Q_{\psi} \left(\frac{1+2j+2i}{4} \right)\sum_{k=1}^{(n-1)/2}2^{2(i+j)}k^{2(i+j)}.
\end{equation}
To establish the correspondence with the large $n$ action 
(\ref{eq:action_at_large_n}), 
we need to trade the sum over $k$ for a function of $n$. 
This is achieved by considering the well-known algebraic condition:
\begin{equation}\label{eq:power_bound}
\frac{n^{p+1}}{p+1} \hspace{0.2cm} < \hspace{0.2cm} \sum_{k=1}^n k^p
\hspace{0.2cm} < \hspace{0.2cm} \frac{(n+1)^{p+1}}{p+1}
\end{equation}
It can be seen that in the limit of large $n$ the lower and upper bounds
converge. Thus, we can make
the following algebraic identification
\begin{equation}
\sum_{k=1}^{(n-1)/2} 2^{2(i+j)}k^{2(i+j)} \sim \frac{(n-1)^{(2j+2i+1)}}{2(1+2j+2i)}.
\end{equation}
Finally, inserting this relation into the action (\ref{actionsofar}),  
collecting together all the terms and substituting for 
$Y$ in terms of $Y_0$, leads us to the result
\begin{equation}
\label{Pnresult}
\fbox{$\displaystyle  P_n \simeq -\frac{T_3 h^4(n-1)^{2j-2i+1}}{4 i!j!} \left\lbrack \left(-\frac{1}{2}\right) \ldots \left(\frac{3-2i}{2}\right)Y_0^i\right\rbrack
\left\lbrack \left(-\frac{1}{2} \right)\ldots \left(\frac{3-2j}{2} \right)(-\dot{\psi}^2)^j\right\rbrack $}
\end{equation}
Eq. (\ref{Pnresult}) may be compared directly to the 
action (\ref{eq:action_at_large_n}) in the large $n$ limit.  

In conclusion, therefore, we have argued that the action for finite $n$ 
does indeed converge to the full action at large $n$, 
which is a non-trivial result. This indicates that the finite $n$ action
can be trusted (at least for the simple case we consider). Whilst the full
structure of the non-Abelian DBI action remains unknown, the Myers 
action \cite{Myers:1999ps}---and therefore our finite $n$ theory given by 
Eq. (\ref{eq:recursive_action})---will be a good approximation to leading
order in an $\alpha'$ expansion.
In the following section, we consider 
some of the consequences of this action for the inflationary scenario 
of the early universe. 


\section{Matrix inflation at finite $n$}\label{sec:finiten}


\subsection{General remarks}

One of the main characteristics of DBI-driven inflation is that the 
inflaton can move relativistically and still drive a sustained period
of accelerated expansion. This is possible because the warped metric  
redshifts all physical scales associated with the brane dynamics. This
feature is frequently exploited to simplify the functional form of 
the solution by considering the so-called `ultra-relativistic limit', where
the kinetic contribution to the action is rewritten in terms of
a function, $\gamma$, corresponding to a generalization 
of the relativistic factor of special relativity.

In our scenario, the algebraic structure of the multi-brane action forces 
us to consider a more general version of this function, which we 
parametrize in terms of the inflaton $\phi$ such that
\begin{equation}
\label{defgammak}
\gamma_k = \left(1-\frac{\dot{\phi}^2}{h^4 T_3} \left(\frac{2k-\delta_n}{n-1} \right)^2 \right)^{-1/2}
\end{equation}
Note that there is an implicit summation over $k$ in this expression. 
Let us study this function on its own for the moment. It is trivial to see 
that increasing the number of branes, $n$, increases the number of 
poles of the function. For example, we see that in the case of $n=7$
\begin{equation}
\gamma_k = \left(1-\frac{\dot{\phi}^2}{h^4 T_3}\right)^{-1/2}+\left(1-\frac{4\dot{\phi}^2}{9 h^4 T_3}\right)^{-1/2}+\left(1-\frac{\dot{\phi}^2}{9h^4 T_3}\right)^{-1/2}
\end{equation}
and it follows that $\gamma$ diverges in the same limit 
as that of the usual DBI-inflation models, namely 
when $\dot{\phi}^2 \sim h^4 T_3$. The divergences arising from
the higher order expansion in $k$ are not physical. 
Consequently, it is the lowest term in $k$ that determines 
the sound speed for inflaton field fluctuations, 
at least if the analysis is limited to the behavior of the 
$\gamma$-function. The implication is that $\gamma_k \sim \gamma$ 
in the ultra-relativistic regime.

If we include the coupling to Einstein-Hilbert gravity, 
we find that the energy density of the DBI-scalar is not conserved, due to the
presence of the FRW scale factor. It is a standard result that 
$\dot{E_n}=-3H(P_n+E_n)$, where $H$ is the usual Hubble factor, and  
from this expression we 
can then write down the dynamics of the Hubble parameter as follows:
\begin{eqnarray} \label{eq:relativistic_continuity}
\dot{H}_n &=& -3H (P+E) \\
&\simeq & 2 T_3 h^4 \sum_{k=1}^{(n-1+\delta_n)/2} \frac{\gamma_k(\gamma_k^2-1)}{\sqrt{1+Y(2k-\delta_n)^2}} \left(3+4Y(2k-\delta_n)^2-\frac{(\gamma_k^2-1)(2+3Y(2k-\delta_n)^2)}{\gamma_k^2} \right) \nonumber
\end{eqnarray}
This is a highly complicated sum over all $k$ states. However, the 
analysis can be simplified somewhat if it is assumed 
that the generalized $\gamma$-function (\ref{defgammak})
is large, i.e. $\gamma_k\gg1$ 
$\forall$ $k$. If this condition is satisfied, we find that 
\begin{equation}
\dot{H}_n \sim - \frac{T_3 h^4 \gamma^3 }{M_p^2} \sum_k \sqrt{1+(2k-\delta_n)^2Y}
\end{equation}
where we have explicitly included the dependence on the number of branes 
in the $\dot{H}$ term and used the fact that $\gamma_k \sim \gamma$ 
in this regime. This expression differs significantly
from the corresponding limit of the single brane ($n=1$) solution. 


\subsection{Non-relativistic expansion}

What happens to the expansion of the action in the non-relativistic limit?
 Related work for the single brane case is discussed in 
\cite{Spalinski:2007kt, Shandera:2004zy}.
Fortunately, one can see that the symmetrized trace operation 
commutes with the non-relativistic limit and therefore
we can immediately write down the 
following equation of motion for the inflaton:
\begin{equation}
\dot{\phi} \simeq -H' M_p^2 \left(\sum_{k=1}^{(n-1+\delta_n)/2} 
\left( \frac{2k-\delta_n}{n-1}\right)^2 
\frac{(3+4Y(2k-\delta_n)^2)}{\sqrt{1+Y(2k-\delta_n)^2}} \right)^{-1}\,.
\end{equation}
Here a prime denotes differentiation with respect to $\phi$ and 
we have assumed the validity of the Hamilton-Jacobi 
expansion when higher order terms in the velocity expansion are neglected. 
Given an expression for the time dependence of the inflaton, 
the corresponding slow-roll parameter $\epsilon \equiv \dot{H}/H^2$ 
can also be deduced: 
\begin{equation}
\epsilon \simeq  M_p^2 \left(\frac{H'}{H} \right)^2 \left(\sum_{k=1}^{(n-1+\delta_n)/2} \left( \frac{2k-\delta_n}{n-1}\right)^2 \frac{(3+4Y(2k-\delta_n)^2)}{\sqrt{1+Y(2k-\delta_n)^2}} \right)^{-1}
\end{equation}
It follows, therefore, that 
the slow-roll parameter is suppressed relative to that of 
canonical inflation by the additional sum in the denominator,
which we will denote by $\sum_k f_k$ for simplicity, i.e.,
\begin{equation}
\epsilon \simeq \frac{M_p^2}{\sum_k f_k} \left( \frac{H'}{H}\right)^2.
\end{equation}
The maximal suppression occurs when $\sum_k f_k$ takes its largest value, 
although one must also be aware that $Y$ is itself a function of $n$, which 
complicates the analysis. Indeed, it is trivial to see that
\begin{equation}
\label{Yphidep}
Y = \frac{4 \phi^4}{h^4 \lambda^2 T_3^2(n-1)^4}
\end{equation}
and so the dependence on the inflaton is essential in determining the dynamics. 

In most cases of interest, the warped throats fall into one of two classes. The first
is an $AdS$-type throat, arising for example from the near horizon limit of $D3$-branes. In this
instance the warp factor takes the form $h \sim \phi$ and $Y$ is fixed at some non-negative value.
The second class corresponds to regularized throats, where the back-reaction of the form fluxes are
used to `cap' the throat at the tip. A canonical example of such a background is given by the
warped deformed conifold \cite{Klebanov:2000hb, Herzog:2001xk}. The warping in this case tends to a constant at some finite value of
the throat, and therefore we find that $Y$ is an increasing (decreasing) function of the inflaton
depending upon whether we are considering small (large) field inflation.
For the $IR$ scenario (an example of small field inflation), 
one sees that $\sum_k f_k$ is rapidly dominated by 
smaller values of $n$ and therefore
inflationary trajectories for $n=2$ are preferred over those with larger $n$. 
In the $UV$ (large field) case, on the other hand, we find that this function 
is sensitive to the precise value of the constants. Consequently, 
we cannot identify a priori whether more suppression occurs for 
smaller values of $n$, although one should remember that 
$UV$ inflation is more severely constrained than the $IR$ scenario.

In general, we can write the Hubble parameter in the following form: 
\begin{equation}
3 H^2 M_p^2 \simeq 2 T_3 \sum_k W_k (1+\varepsilon_k)
\end{equation}
where the generalized effective potential is given by 
\begin{equation}
W_k \simeq h^4(1+2Y(2k-\delta_n)^2)+\left(\frac{n}{n-1+\delta_n}\right)(V-h^4)\sqrt{1+Y(2k-\delta_n)^2}
\end{equation}     
and the parameter
\begin{equation}\label{eq:epsilonk}
\varepsilon_k \simeq \frac{\dot{\phi}^2}{2 T_3 (n-1)^2 W_k}\frac{(3+4Y(2k-\delta_n)^2)}{\sqrt{1+Y(2k-\delta_n)^2}}
\end{equation}
quantifies the ratio of the kinetic and potential energies. 
It follows, therefore, that the generalized master (Friedmann) 
equation for the Hubble parameter can be expressed in the form 
\begin{equation}
3H^2 M_p^2 -2 T_3 \sum_k W_k - \sum_k
\frac{(3+4Y(2k-\delta_n)^2}{\sqrt{1+Y(2k-\delta_n)^2}} \frac{H'^2
M_p^4}{(n-1)^2} \left(\frac{1}{\sum_p f_p}\right)^2 \simeq 0 
\end{equation}
and this equation can be employed to determine the inflationary trajectories.


\subsection{A matrix inflation model in AdS}

In this subsection we investigate a concrete example of 
non-relativistic matrix inflation with the aim of identifying 
regions of parameter space that are consistent with cosmological 
observations. 
Specifically, we consider IR (small field) inflation driven by an inverted 
harmonic oscillator potential, where the branes are propagating out of 
an AdS throat. 

It is simple to show by expanding Eq. (\ref{eq:recursive_action}) that the 
leading order expression for the pressure can be written as
\begin{equation}
P_n = Q \dot{\phi}^2 - nT_3 V(\phi)+T_3 h^4 \left(n-\sum_k \frac{2+4(2k-\delta_n)^2Y}{\sqrt{1+(2k-\delta_n)^2Y}} \right)
\end{equation}
where we have defined 
\begin{equation}
\label{defQ}
Q \equiv \sum_k \frac{(3+4(2k-\delta_n)^2Y)(2k-\delta_n)^2}{(n-1)^2\sqrt{1+(2k-\delta_n)^2Y}}.
\end{equation}
For a class of $AdS$ backgrounds, the warp factor 
$h(\phi ) = \phi/(L\sqrt{T_3})$
is determined in terms of the AdS radius of curvature 
$L=[4\pi^4g_sMK/({\rm Vol}(X_5)m_s^4)]^{1/4}$, where
$g_s$ denotes the string coupling, ${\rm{Vol}(X_5)}$ is the volume of the 
base and the background fluxes $M,K$ thread through the three-cycles 
in the compactification. Eq. (\ref{Yphidep}) then  
implies that $Q$ is independent of $\phi$. 
We may therefore introduce a new field, $\psi = \sqrt{2 Q} \phi$,
in order to write the pressure in a canonical form. The effective potential 
energy of this field is then given by
\begin{equation}
W = n T_3 V(\psi)-\frac{ \rm{Vol}(X_5) \psi^4}{2 \pi M K Q^2}\left(n-\sum_k \frac{(2+4(2k-\delta_n)^2Y)}{\sqrt{1+(2k-\delta_n)^2Y}} \right)
\end{equation}
where we have written the explicit parameters
\begin{equation}
\label{defY_0}
Y = \frac{Y_0}{(n-1)^4}, \hspace{1cm} 
Y_0 = \frac{4 \pi^2 g_s MK}{\rm{Vol}(X_5)}
\end{equation}
Since typically the product of fluxes must satisfy $MK\gg1$, it follows  
that the second term in the effective potential is 
suppressed and therefore only the term proportional to 
$V(\psi)$ is expected to dominate the dynamics. 

Indeed, if one assumes that the inflaton potential 
takes the form of a simple inverted harmonic oscillator, 
$V(\phi ) = nT_3V_0 (1-\beta \phi^2 /M_p^2)$, 
where $\beta$ is an order one parameter of the theory, we may write
\begin{equation}
\label{hilltop}
W (\psi ) =W_0 - \frac{\omega^2 \psi^2}{2}
\end{equation}
where 
\begin{equation}
W_0 = nT_3 V_0, \hspace{1cm} \omega^2 = \frac{\beta n T_3 V_0}{Q M_p^2}\,.
\end{equation}
Coincident branes located 
near the tip of a warped throat are expected to be attracted towards 
branes residing in other throats through the generation of such a tachyonic 
potential. 
Since our theory is now in canonical form, we can adapt 
the usual inflationary tools to our
current objective. During inflation the Gaussian curvature 
perturbation has an amplitude given by
\begin{equation}
\mathcal{P_R} = \frac{W}{24 \pi^2 M_p^4 \epsilon}
\end{equation}
and spectral index
\begin{equation}
1-n_s = 6\epsilon-2\eta
\end{equation}
where $\epsilon \equiv  
\frac{1}{2} M_p^2(W'/W)^2$ and $\eta \equiv M_p^2W''/W$
are the usual slow roll parameters obtained from derivatives 
of the effective potential. Near the maximum of the potential we see that
\begin{equation}
\eta \simeq -\frac{M_p^2 \omega^2}{W_0}, 
\hspace{0.5cm} \epsilon \simeq \frac{M_p^2 \omega^4 \psi^2}{2 W_0^2} 
\sim \frac{\eta^2 \psi^2}{2 M_p^2}
\end{equation}
which implies that $\epsilon \ll  |\eta| \ll 1$. The spectral index 
of the density perturbations can therefore be computed as
\begin{equation}\label{eq:def_Q}
1-n_s = \frac{2\beta}{Q}
\end{equation}
This has an interesting dependence on the number of coincident branes, 
since $Q$ is not monotonic. Indeed, $Q$ has a local maximum when $n$ 
takes the lowest possible value ($n=2$ for the even case), 
which suggests that $1-n_s$ is initially small. 
As we increase the number of branes, we see that $1-n_s$ also 
increases until it attains a local maximum around $n\sim6$ (for the even case),
before asymptotically tending to zero as $n$ increases further.

The above expressions assume implicitly that the slow-roll
conditions hold until the very end of inflation, i.e., 
it is assumed that the branes continue to move non-relativistically. 
The self-consistency of this assumption can be verified  
by employing the effective field equation $3H \dot{\psi} \simeq 
- dW/d\psi$ and the Friedmann equation $3H^2 \simeq W_0/M_p^2$, together
with Eq. (\ref{eq:def_Q}).  The non-relativistic 
limit, $\dot{\phi}^2 \ll T_3 h^4$, is then equivalent to  
\begin{equation}
\label{nonrel}
\psi^2 \gg \beta L^4 T_3(1-n_s)\frac{n T_3 V_0}{3 M_p^2}
\end{equation}
Since the inflaton is a monotonically increasing function in this 
scenario, it suffices to show that the bound (\ref{nonrel}) was satisfied 
when observable scales crossed the Hubble radius. 
The value of the field
at that time is related to the normalization of the CMB power spectrum 
such that 
\begin{equation}
\label{COBEnorm}
\frac{\psi_{\rm cmb}^2}{M_p^2} = \frac{1}{3 \pi^2(1-n_s)^2 \mathcal{P_R}}
\frac{nT_3 V_0}{M_p^4}
\end{equation}
and substituting Eq. (\ref{COBEnorm}) into the constraint 
(\ref{nonrel}) then yields the consistency condition
\begin{equation}
\beta T_3 L^4 (1-n_s)^3 \pi^2 \mathcal{P_R} \ll 1
\end{equation}
This constraint can typically be satisfied for a large 
region of the physical parameter space.

The value of $\psi_{\rm cmb}$ can also be determined from the 
scalar field equation of motion to be
\begin{equation}
\psi_{\rm cmb} = \psi_{\rm end}e^{-(1-n_s)N/2}
\end{equation}
This allows us to write the tensor-scalar ratio in the following form
\begin{equation}
r = 16 \epsilon =  2(1-n_s)^2 \frac{\psi_{\rm cmb}^2}{M_p^2}
\end{equation}
suggesting that gravitational waves are negligible in this model.

We may gain further insight by specializing to the case of even or odd $n$. 
Let us choose the latter case for clarity and reconsider 
the constraint on the spectral index (\ref{eq:def_Q}). 
We can clearly see that since $Y$ is positive-definite,
the definition of $Q$ given in Eq. (\ref{defQ}) 
implies that this parameter is bounded from above
such that 
\begin{equation}
Q < \sum_{k=1}^{(n-1)/2} \left( \frac{12 k^2}{(n-1)^2} + \frac{64 k^4 Y_0}{(n-1)^6}\right).
\end{equation}
Utilizing the bound (\ref{eq:power_bound}) enables us 
to write down a further constraint 
that is slightly weaker, but expressible simply as a function of $n$:
\begin{equation}\label{eq:Q_bound}
Q < \frac{12}{3(n-1)^2}\left(\frac{n+1}{2} \right)^3+ \frac{64 Y_0}{5 (n-1)^6}\left(\frac{n+1}{2} \right)^5 + \ldots
\end{equation}
where we neglect the terms coming from higher orders in the expansion. 

The WMAP five-year data implies that $n_s \ge 0.93$ $(2\sigma)$
when the gravitational wave background is negligible. This 
in turn imposes the condition 
$Q \ge 30 \beta \gg n$ if we take $\beta \simeq \mathcal{O}(1)$. Now, 
since the first term in the expansion of (\ref{eq:Q_bound}) is essentially
$\mathcal{O}(n)$, and therefore negligible compared to $Q$,  
the leading order bound is dominated by
\begin{equation}
Q \le \frac{64 Y_0}{5 (n-1)^6} \left(\frac{n+1}{2} \right)^5
\end{equation}
and this results in $Y_0$ being bounded from below:
\begin{equation}
\label{Y_0condition}
Y_0 \ge \frac{5 \beta}{(1-n_s)} \frac{(n-1)^6}{(n+1)^5}
\end{equation}
In principle, therefore, condition (\ref{Y_0condition}) 
may be interpreted as 
an observational lower bound on the product of the fluxes $MK$ 
after `canonical' values 
for the string coupling, $g_s \simeq 10^{-2}$, and base manifold volume, 
${\rm Vol}(X_5) \simeq \pi^3$, have been specified.  It follows 
from Eq. (\ref{defY_0}) that 
\begin{equation}
\label{MKbound}
MK \ge \frac{400 \beta}{(1-n_s)} \frac{(n-1)^6}{(n+1)^5}\,.
\end{equation}
Moreover, the ratio $(n-1)^6/(n+1)^5$ is a monotonically 
increasing function of $n$ and this implies that consistency 
with observations would be difficult to achieve if condition 
(\ref{MKbound}) was not satisfied for $n=3$. For example, 
invoking the central value 
for the spectral index inferred from WMAP, $n_s = 0.96$, and setting 
$\beta \simeq  \mathcal{O} (1)$ suggests that $MK > 630$. 
One should note that the fluxes depend on the inverses of 
both the string coupling and the spectral index, and
a more weakly coupled theory therefore 
leads to a tighter bound on the fluxes. Likewise,
it becomes more difficult to satisfy this condition as the spectrum 
becomes closer to scale invariance. 

To summarize, in this model the spectral index of the 
density perturbations is determined by the parameter $Q$ that quantifies 
the non-Abelian nature of the multi-brane configuration. 
Consistency with observations is possible if the 
background fluxes are sufficiently large. 

In the following section, we proceed to investigate the dynamics 
of the $n=2$ configuration in more detail. 


\section{Cosmology of the two-dimensional representation}\label{sec:twodim}

Given the recursive structure of the action in Eq. (\ref{eq:recursive_action}), 
we see that the two dimensional representation\footnote{Note that this 
is the adjoint representation of the group.} 
of $SU(2)$ is the most important,
corresponding in our matrix language to a configuration of 
two coincident $D3$-branes. 
Note that this is not the same model as that considered in \cite{Cai:2009hw}, 
which considered two $D3$-branes separated by a distance larger 
than the string length. The world-volume gauge theory in that case 
is therefore $U(1)\times U(1)$, whereas
our model possesses a $U(2)$ symmetry. 
Note that this is the gauge group of the open string states on the world-volume, and should not
be confused with the $SO(3) \sim SU(2)$ symmetry group that parametrizes the transverse space.


\subsection{The relativistic limit}

To begin this section let us focus on the relativistic limit to highlight 
how the theory differs from the $n=1$ solution.
Unlike the case of a single $D3$-brane, the pressure and energy of the 
model have very different dependence upon the brane velocity. 
Given our generalized definition of $\gamma_k$, it is convenient to now 
introduce the following notation
\begin{equation}
\gamma_2 = \frac{1}{\sqrt{1-Z\dot{R}^2}}\,.
\end{equation}
As in the case of $n=1$ the sound speed of the Fourier modes, $c_s$, is 
reduced from unity---the relevant calculation proceeds in the 
usual manner and we find that
\begin{equation}
c_s^2 = \frac{1}{2 \gamma_2^2} \left(\frac{2\gamma_2^2(3+4Y)-(\gamma_2^2-1)(4+Y)}{\gamma_2^2(3+4Y)-(\gamma_2^2-1)Y} \right)
\end{equation}
which is very different from that of the standard
DBI inflation models (see \cite{Peiris:2007gz} for example). 
In the ultra-relativistic limit the above expression simplifies to become
\begin{equation}
c_s^2 \simeq \frac{1}{6 \gamma_2^2} \left(\frac{2+7Y}{1+Y} \right)
\end{equation}
and this should be compared to the $c_s^2 \sim 1/\gamma^2$ dependence of  
the $n=1$ model. One sees that for $Y\gg1$ the sound speed varies as 
$(7/6) \gamma_2^{-2}$, whilst
in the converse limit it behaves as $(1/3) \gamma_2^{-2}$. 
More importantly for cosmology, the sound speed modifies
the tensor-scalar ratio as follows:
\begin{equation}
r \simeq \frac{16 \epsilon_H}{\gamma_2} \sqrt{\frac{1}{6} \left(\frac{2+7Y}{1+Y}\right)},
\end{equation}
with the presence of an additional enhancement factor due to the 
contribution of $Y$. Indeed, this accounts for the enhancement of $r$, 
relative to single brane models, discussed in \cite{Huston:2008ku}.

Given the modified expression for the sound speed, and the non-linear 
form of the action, we can compute the bi-spectrum 
arising from higher order interactions of the different Fourier modes in the 
density perturbations. For simplicity, we focus on the equilateral triangle
configuration to highlight the relative contribution 
to the non-Gaussian parameter $f_{nl}$. A standard calculation reveals that 
\begin{eqnarray}
f_{nl} &\sim& -\frac{85 \gamma_2^2}{162}\left(\frac{\gamma_2^2(3+4Y)-Y(\gamma_2^2-1)}{2\gamma_2^2(3+4Y)-(\gamma_2^2-1)(4+Y)} \right)+\frac{85}{324} \nonumber \\
&-& \frac{5(\gamma_2^2-1)}{81}\left( \frac{\gamma_2^2(5+6Y)-Y(\gamma_2^2-1)}{\gamma_2^2(3+4Y)-Y(\gamma_2^2-2)}\right).
\end{eqnarray}
This function is negative definite for all values of 
$\gamma_2$ and $Y$ in the relativistic limit. Moreover, one sees that
the detectable range of $f_{nl}$ falls into the regime where $Y\ll1$. For example, 
in $AdS$-type backgrounds, where $h\sim \phi$, the parameter 
$Y$ reduces to a constant determined by Eq. (\ref{defY_0}). 
For sufficiently large $MK/g_s$,  
we find that $Y$ is suppressed, and detectable levels of 
(negative) $f_{nl}$ are therefore more likely.
Note that in the limit where $Y \to 0$, the functional form of the 
action becomes very similar to that of the single brane models.  
Thus, a small (but manifestly non-zero) value of $Y$ is 
crucial for generating a detectable signature of non-Gaussianity
in the primordial curvature perturbation.


Although the highly non-linear nature of the 
multi-brane action leads to interesting observables, it is also 
of interest to investigate the nature of the inflationary trajectories.  
This can be achieved by calculating the variation of the Hubble 
parameter from Eq. (\ref{eq:relativistic_continuity}), which takes the form
\begin{equation}\label{eq:n=2hdot}
\dot{H} \sim -\frac{T_3 h^4 \gamma_2 (\gamma_2^2-1)}{M_p^2}\frac{1}{\sqrt{1+Y}}\left(3+4Y+\frac{(2+3Y)(\gamma_2^2-1)}{\gamma_2^2} \right).
\end{equation}
After taking the ultra-relativistic limit of Eq. (\ref{eq:n=2hdot}), 
as well as the corresponding limit in the definition of the energy density
\begin{equation}
E_2 \sim 2 T_3\left(h^4 \gamma_2^3\sqrt{1+Y} + V - h^4 \right)
\end{equation}
we see that the `slow roll' parameter, $\epsilon =\dot{H}/H^2$, becomes
\begin{equation}
\epsilon \sim \frac{3}{2} \left(1+\frac{(V-h^4)}{h^4 \gamma_2^3\sqrt{1+Y}} \right)^{-1}\left(1+ \frac{1}{\gamma_2^2} \frac{(1+2Y)}{(1+Y)}\right)
\end{equation}
This leads to a novel phenomenon in multi-brane DBI models. 
The final term in parenthesis is essentially 
a term of order one suppressed by the relativistic factor and it 
effectively decouples in the relativistic limit.
If the effective potential $(V-h^4)$ is much smaller than the kinetic term, 
inflation will never begin. On the other hand, 
if the potential term dominates---even for the fast rolling
solution---we see that this is a concrete realization of eternal
inflation. 


\subsection{The non-relativistic limit}

We now proceed to consider the non-relativistic limit of the theory. 
In this limit, the energy density takes the following, almost canonical, form:
\begin{equation}
E_2 = \frac{2 T_3}{\sqrt{1+Y}} \left( W(\phi)+\frac{\dot{\phi^2}}{2T_3}(3+4Y)\right)
\end{equation}
where the effective potential is given by 
\begin{equation}
W(\phi) = h^4(1+2Y-\sqrt{1+Y})+V(\phi)\sqrt{1+Y}.
\end{equation}
We also note that the pressure, $P_2$, in the non-relativistic limit 
is given by
\begin{equation}
P_2(Z,Y)= - \frac{2 T_3}{\sqrt{1+Y}} \left(W(\phi) - \frac{ \dot{\phi}^2}{2T_3}(3+4Y) \right).
\end{equation}
As an aside, we briefly comment on the non-Gaussianities in this limit.
In the $n=1$ case, the non-relativistic limit 
reduces to a canonical scalar
field theory, and therefore one expects that 
$f_{nl}$ will be of the same order as the slow roll parameters. 
For $n=2$, it is convenient to 
parametrize the non-relativistic expansion as 
$\gamma_2 \sim 1 + \xi$ and focus on the correction term
\begin{equation}
f_{nl} \sim - \frac{5 \xi}{234} \left(\frac{210+167Y}{3+4Y} \right)
\end{equation}
One can see that this term is very small even in the limit 
where $Y\gg1$.

Assuming the validity of the Hamilton-Jacobi approximation, 
we can use the continuity equation to 
obtain the following expression for the time-dependence of the scalar field
\begin{equation}
\dot{\phi} = -M_p^2 H'\frac{\sqrt{1+Y}}{3+4Y}\,,
\end{equation}
which is modified from that of the canonical theory by the $Y$-dependent terms. 
Interestingly, we can write the slow roll parameter $\epsilon$ in
in exactly the same way as the canonical model 
\begin{equation}
\epsilon = \frac{3 \epsilon_2}{1+\epsilon_2} 
\end{equation}
so that $\epsilon_2 \le 1/2$ is required in order for inflation to occur. 
Recall that $\epsilon_2$ is the ratio of the
kinetic energy to the potential energy in the theory and follows from 
the generalized definition in (\ref{eq:epsilonk}).

The final step for our analysis is to calculate the master 
(Friedmann) equation for the Hubble parameter. This is given by 
\begin{equation}
 3 M_p^2 H^2 \frac{3+4Y}{\sqrt{1+Y}}-2T_3 W \left(\frac{3+4Y}{1+Y}\right) = M_p^4 H'^2
\end{equation}
which follows from the generalized expressions derived 
in section 3.2. Since an inflating solution requires that the terms on the 
right hand side are smaller than those on the left hand side, 
we can iteratively expand the Hubble parameter in a derivative series. 
The decoupled leading order solutions are given by 
\begin{eqnarray}
0 &\sim& 3M_p^2 H_0^2-2WT_3 \\
M_p^2 H_0^{'2} &\sim& 6 H_0 H_1 \frac{(3+4Y)}{\sqrt{1+Y}} \nonumber \\
2M_p^2 H_0'H_1' &\sim& (2H_0H_2+H_1^2)\frac{(3+4Y)}{\sqrt{1+Y}} \nonumber \\
M_p^2(H_1'^2+2H_0'H_2') &\sim& 6H_1H_2 \frac{(3+4Y)}{\sqrt{1+Y}} \nonumber
\end{eqnarray}
for the first few terms in the expansion and this enables us to 
reconstruct the Hubble parameter at leading order as follows:
\begin{equation}
H \sim \sqrt{\frac{2 T_3 W}{3M_p^2}} \left(1+\frac{1}{12}\sqrt{\frac{3(1+Y)}{2T_3 W}} \frac{M_p^3 W'}{W(3+4Y)}+\ldots \right)\,.
\end{equation}
The validity of this expression requires that the second order terms 
are negligible in comparison, which is equivalent to the condition
\begin{equation}
1\gg \frac{3 M_p^2}{2} \frac{\sqrt{1+Y}}{(3+4Y)} \left(\frac{W''}{W}-\frac{W'^2}{W^2}- \frac{W'Y'}{2W}\frac{(5+4Y)}{(1+Y)(3+4Y)}-\frac{M_pW'}{12W}
\sqrt{\frac{2}{3 T_3 W}} \right).
\end{equation}
One can see that in the asymptotic regimes for $Y$, 
this constraint is essentially a bound on the values of $W$ and its derivatives, 
exactly as in canonical slow roll inflation. This is particularly 
clear for the case where $Y$ is constant, since this implies that the effective 
potential depends solely upon the
derivatives of the warp factor $h(\phi)$ and the scalar potential $V(\phi)$.
The limit where $Y \to 0$ is essentially that of the usual DBI scenario 
(modulo a field redefinition) and we can therefore 
be sure that inflation will occur. 
In the limit of large $Y$,  we see that the higher-order 
corrections to the Hubble parameter become arranged 
into a $1/Y$ expansion and are therefore suppressed 
relative to the leading order terms.

\subsection{Dynamical trajectories}

In this section, we are interested in studying the flow equations for
the $n=2$ solution along the research-lines of 
\cite{Tzirakis:2008qy, Kinney:2007ag, Peiris:2007gz}
in both the relativistic and non-relativistic
limits. Recall that in the case of a single brane, 
the inflaton equation of motion
(in the Hamilton-Jacobi formalism) is given by 
\begin{equation}\label{eq:n=1eom}
\dot{\phi} \simeq - \frac{2 M_p^2 H'}{\gamma}
\end{equation}
and the first slow roll parameter $\epsilon_H$ by 
\begin{equation}
\epsilon_H \simeq \frac{2 M_p^2}{\gamma}\left(\frac{H'}{H} \right)^2. 
\end{equation}
In the case of finite $n$, the non-linear 
(and occasionally higher derivative) nature of the action 
prevents us from analyzing the flow equations
in the usual manner. However, there should exist regions 
of parameter space where one can work with equivalent algebraic 
structures. As such, we may extend the relations first 
established in \cite{Peiris:2007gz} by introducing the following
generalized flow parameters:  
\begin{eqnarray}
\label{flow}
\epsilon(\phi) &\sim& \frac{2 M_p^2}{f(\phi)} \left(\frac{H'}{H}\right)^2 \nonumber \\
\eta(\phi) &\sim& \frac{2 M_p^2}{f(\phi)} \frac{H''}{H} \nonumber \\
\kappa(\phi) &\sim& \frac{2 M_p^2}{f(\phi)}\frac{H'}{H} \frac{f(\phi)'}{f(\phi)} \\
{}^l \lambda &\sim& \left(\frac{2 M_p^2}{f(\phi)} \right)^l \left(\frac{H'}{H} \right)^{l-1} \frac{1}{H} \frac{d^{l+1}H}{d\phi^{l+1}}  \nonumber \\
{}^l \alpha &\sim& \left(\frac{2 M_p^2}{f(\phi)} \right)^l\left(\frac{H'}{H} \right)^{l-1} \frac{1}{f(\phi)} \frac{d^{l+1}f(\phi)}{d^{l+1}\phi} \nonumber
\end{eqnarray}
where $f(\phi)$ represents a generalized function that 
needs to be determined in a limit of the theory where 
higher derivative terms can indeed be neglected\footnote{The flow 
parameters used in this context are only applicable to the leading 
order dimension four terms. One can include the higher derivative
operators, but the flow equations cannot then be written 
as simple functions of $(H'/H)$. We hope to return to this more 
general problem in a future publication.}.

It is convenient for our discussion 
to note the following relations between the parameters:
\begin{equation}
\eta ={}^1 \lambda, \hspace{0.5cm} \xi = {}^2 \lambda, \hspace{0.5cm} \rho = {}^1 \alpha, \hspace{0.5cm} \sigma = {}^2\alpha\,.
\end{equation}
It then follows that 
the dynamics of these parameters can be established by solving the set of differential equations 
\begin{eqnarray}
\frac{d \epsilon}{d N_e} &\sim& -\epsilon(2\epsilon-2\eta+\kappa) \nonumber \\
\frac{d \eta}{d N_e} &\sim& -\eta(\epsilon+\kappa) + \xi \nonumber \\
\frac{d \kappa}{d N_e} &\sim& -\kappa(2\kappa+\epsilon-\eta)+\epsilon \rho \\
\frac{d \rho}{d N_e} &\sim& -2 \rho \kappa + \sigma \nonumber \\
\frac{d^l \lambda}{dN_e} &\sim& -{}^l\lambda(l\kappa+l\epsilon-\eta(l-1))+{}^{l+1}\lambda \nonumber \\
\frac{d^l \alpha}{d N_e} &\sim& -{}^l\alpha ((l+1)\kappa + (l-1)\epsilon -
(l-1)\eta)+{}^{l+1}\alpha \nonumber ,
\end{eqnarray}
where $N_e$ represents the number of e-foldings \emph{before} the end of inflation and is defined  
by the condition $d N_e = - H dt$. 
The critical (stable) points of the phase space trajectories 
then correspond to the vanishing of each term on the right hand sides.

A simple calculation shows that at leading order in the derivative
expansion, the function $f$ for the $n=2$ case becomes
\begin{equation}
f \sim \frac{2 \gamma_2^3}{\sqrt{1+Y}}\,.
\end{equation}
This is significantly different to the analogous term for the single brane 
model, where $f \sim \gamma$. Moreover, 
since $Y$ can also depend on the field $\phi$ 
(provided $h$ is constant), we find that there is an 
interesting scenario where the branes are moving relativistically, 
but $f \sim {\rm  constant}$. 
This feature will be important in understanding the 
full dynamical phase space flow. 
In the non-relativistic limit we can truncate the theory 
such that $\gamma_2 \sim 1$, and then we find that 
$f$ is only a function of $Y$. In this case, $f$   
can be constant if the warping is of the AdS form $h \sim \phi$. 
Thus there are two solution branches that are of interest if we wish $f$ to be constant:
\begin{itemize}
\item Relativistic limit and taking $h \sim $ constant.
\item Non-relativistic limit and taking $h \sim \phi$.
\end{itemize}
Note that the latter branch leads to the strongest constraint, 
since $f$ is always constant if the warping is of the AdS-type.
Indeed, in this case we immediately see that 
$\kappa = 0 = {}^l \alpha$ which reduces the 
dimension of the overall phase space.
Since the flow parameter $\epsilon$ is first order, 
we take this to be the fundamental parameter for our analysis and 
outline the results below: 
\begin{itemize}
\item $\epsilon = 0$ 

Setting $\epsilon$ to be zero fixes $H'$ to be zero and therefore 
the scale factor can be trivially integrated to yield
\begin{equation}
a(t) \sim e^{Ht}
\end{equation}

\item $\epsilon \ne 0$

Here we are forced to choose solution slices where $\eta  = \epsilon$ and 
also ${}^l \lambda = \epsilon^l$, as can be seen from the flow equations. Again
from this we can infer that $H$ is fixed to be exponential in form
\begin{equation}
H \sim H_0 \exp \left(\pm \sqrt{\frac{\epsilon f}{2 M_p^2}}\phi\right)
\end{equation}
where we must take the positive sign above and also assume that $\phi$ is 
a decreasing function if this is to lead to an inflationary trajectory. 
From this we can
then reconstruct the solution for the inflaton field velocity with the 
result that $Ht \sim 1/\epsilon$. Therefore we obtain the usual power law 
trajectory with $a \sim t^{1/\epsilon}$ as in the literature 
\cite{Kinney:2007ag, Spalinski:2007qy, Spalinski:2007dv}.
\end{itemize}
These are the only possible behaviors in this regime. 
The final case to consider is when $f$ is no longer constant 
but scales in such a way such that  $\epsilon, \kappa$ remain constant.
In this instance we can solve most of the dynamical equations exactly, 
obtaining relations such as
\begin{equation}
\eta  = \frac{1}{2} (2 \epsilon+\kappa).
\end{equation}
The key point is that one can reconstruct $f$ in terms of the flow 
parameters. It will be convenient to perform the analysis on a case 
by case basis.
Let us first consider the non-relativistic regime of the theory, 
where $f \sim 2/ \sqrt{1+Y}$. We note immediately that $h \sim \phi$ 
yields constant $Y$ and
therefore constant $f$, forcing us to consider the situation where 
$h \sim $ constant. We now have two choices depending upon whether we 
wish to consider
small or large field inflation. In the former case 
it can be seen that $f \to 0$ 
as the branes move and therefore we obtain
\begin{equation}
f_{IR} \sim 8 M_p^2 \epsilon f_{*} \frac{(\sqrt{f_{*}}\kappa \phi \pm \sqrt{8 M_p \epsilon})^2}{(f_* \kappa^2 \phi^2-8 M_p^2 \epsilon)^2} 
\end{equation}
where $f_{*}$ is a constant of integration. It can be verified  
that $f \to 0$ as $\phi$ diverges. For large field inflation, 
we can write the function as follows: 
\begin{equation}
f_{UV} \sim 16 M_p^2 \epsilon \frac{(\sqrt{2}\kappa \phi \pm \sqrt{8 M_p \epsilon})^2}{(2 \kappa^2 \phi^2-8 M_p^2 \epsilon)^2}
\end{equation}
which satisfies the required boundary conditions. The Hubble parameter 
can then be reconstructed trivially (in both cases) to yield 
\begin{equation}
H \sim H_0 \exp \left( \mp \sqrt{\frac{\epsilon}{2 M_p^2}} \int \sqrt{f} d\phi \right)
\end{equation}
Since the resulting expressions are rather complicated,  
we do not write them explicitly. We note, however, that there are many 
different choices for inflating trajectories 
depending on the particular choice of constants (and their signs).

The relativistic case is more complicated, since $f$ can still remain 
constant if the various parameters scale in the right way. 
For simplicity, let us consider
the case where $Y$ is constant and we are also in the large field branch of solution 
space. This requires us to consider a limit where $\gamma_2$ diverges when $\phi \to 0$ in such a way
that we can re-arrange the equation to solve for the relativistic factor as a function of the inflaton
\begin{equation}
\gamma^3 \sim \frac{2 M_p^2 \epsilon (1+Y)}{9 \kappa^2 \phi^2}\,.
\end{equation}
This yields the result that $\phi^2 \propto \exp(-3\kappa N)$
and implies that the field falls off significantly faster 
than in standard single brane inflation. After substituting this
dependence into the definition of 
the flow equation for $\epsilon$, we find that 
\begin{equation}
H \propto \phi^{-\xi}, \hspace{0.5cm} \xi = \frac{\epsilon}{\kappa} \sqrt{\frac{2(1+Y)}{9}}
\end{equation}
In this case, therefore, the inflationary dynamics are 
sensitive to the ratio $\epsilon/\kappa$, as in the single brane case, 
but also to the precise value of $Y$ which is, in turn, set by the
value of the background throat charge. 


\section{Discussion}\label{sec:discuss}

In this work we have considered the cosmological consequences 
of using the action for $n$ coincident $D3$-branes in the finite $n$ limit. 
We have argued that the action for $n$ coincident $D3$-branes, with 
$n \gg  1$ and with scalars 
transforming under the $n$ dimensional representation of $SO(3)$, 
is the same as that arising from a single $D5$-brane wrapping 
an $S^2$ and carrying $n$ units of $U(1)$ flux.
Through the usual notion of string duality, this implies that we 
can either consider cosmology on $n$ coincident $D3$-branes, 
or a single wrapped $D5$-brane
with commutative flux.

One immediate consequence is that the probe limit may no 
longer be a valid description of the physics. On the macroscopic side, 
this is due to the addition
of $U(1)$ flux through the $S^2$, which effectively contributes a 
large  mass correction to the theory at large $n$. Therefore, 
it is preferable to consider
theories with finite $n$ in order to minimize the effect of back-reaction.
This is a non-trivial problem in general, with the only conjectured 
solution being that of the $SO(3)$ scalar representation, 
which has been shown to lead to interesting physics in its own right.

Given that the finite $n$ prescription and the large $n$ prescription 
are essentially decoupled from one another, we have endeavored 
in section 2 to show that
the finite $n$ action does indeed converge to that at large $n$ 
in a certain limit. This is the first time that such an argument has been 
presented in the literature. How does one then interpret the finite $n$ 
theory from the macroscopic side? Although we have not attempted to answer 
such a question, we
believe that the correct interpretation is that of a wrapped $D5$-brane 
which now carries non-commutative $U(1)$ flux on the world-volume through
the introduction of a star product. Thus our $D3$-brane model should be 
dual to a class of non-commutative field theories which have been explored 
in the literature \cite{Seiberg:1999vs, Douglas:2001ba}. It would be 
interesting to explore this in more detail
since this is a different theory to the $\kappa$-deformed models and therefore
evades the torsional interpretation \cite{Gibbons:2009af} of the Snyder 
algebra proposed in \cite{Snyder:1946qz},
which seems to rule out many non-commutative field theory models.

The above discussion applies for the relativistic theory, 
and one may be interested in the preservation (or not) 
of non-Abelian physics in the non-relativistic expansion. 
This was explored in sections 3 and 4 of the paper. In section 3,  
we considered various consequences for the inflationary scenario  
when an arbitrary number of coincident branes are present
and demonstrated explicitly 
how the slow roll parameters are modified by the non-Abelian structure. 
We also considered a particular small-field 
inflationary model where the branes are moving out of 
an AdS-type throat. We found that observational 
bounds on the scalar spectral index can in turn 
constrain the fluxes in the supergravity theory.

In section 4, we discussed  
the case of $n=2$ in an arbitrary background in both 
the relativistic and non-relativistic limits. In 
the latter case, we found that the level of non-Gaussianity in the 
primordial curvature perturbation is  
indeed suppressed, although corrected from the canonical slow roll models. 
We concluded by considering the flow trajectories
for such a configuration and compared these to those 
of the standard, single brane DBI scenario.

We have not addressed the question of reheating in this 
scenario and this is an important topic to consider. 
Neither have we attempted to identify
the location of the standard model. On the other hand, we have argued 
from the field theory perspective
that a $U(n)$ field-theory with scalars transforming under $SO(3)_n$ is 
dual to a $U(1)$ field-theory with scalars transforming under another $U(1)$. 
This duality appears to be true \emph{for all n}. It is tempting to 
suggest that one could then identify standard model-type 
states within a $U(1)$ theory (with additional gauge fields), 
after symmetry breaking. This could be a 
particularly useful description for transferring inflationary 
energy into the particle sector.

\begin{center}
{\bf Acknowledgements}
\end{center}
We wish to thank Andrei Frolov and Steve Thomas for their comments. AB and JW are supported in part by NSERC of Canada.

\end{document}